\documentclass[doublecol]{epl2}

\usepackage{amsfonts,amssymb,amsmath,graphicx,epsfig,amscd}
\usepackage{array}

\def\bi{\begin{itemize}}

\def\ei{\end{itemize}}
\def\be{\begin{equation}}
\def\ee{\end{equation}}
\def\bea{\begin{eqnarray}}
\def\eea{\end{eqnarray}}

\def\gdot{\dot\gamma}

\DeclareTextSymbol{\degre}{OT1}{23}

\title{Investigation of shear banding in three-dimensional foams}

\author{G. Ovarlez\inst{1} \and K. Krishan\inst{2,3} \and S. Cohen-Addad\inst{3}}
\shortauthor{G. Ovarlez \etal}

\institute{\inst{1} Universit\'e Paris-Est, Laboratoire Navier
(UMR 8205 ENPC/LCPC/CNRS), 2 All\'ee Kepler, 77420 Champs-sur-Marne, France\\
\inst{2}School of Physical Sciences, Jawaharlal Nehru University,
New Delhi - 110067, India\\
\inst{3} Universit\'e Paris-Est, Laboratoire de Physique des
Mat\'eriaux Divis\'es et des Interfaces (FRE3300 CNRS), 5 Bd
Descartes, Champs-sur-Marne, 77454 Marne-la-Vall\'ee C\'edex 2,
France} 

\abstract{We study the steady flow properties of different
three-dimensional aqueous foams in a wide gap Couette geometry.
From local velocity measurements through Magnetic Resonance
Imaging techniques and from viscosity bifurcation experiments, we
find that these foams do not exhibit any observable signature of
shear banding. This contrasts with two previous results
(\Name{Rodts \etal} \REVIEW{Europhys. Lett.}{69}{2005}{636} and
\Name{Da Cruz \etal} \REVIEW{Phys. Rev. E}{66}{2002}{051305}); we
discuss possible reasons for this dicrepancy. Moreover, the foams
we studied undergo steady flow for shear rates well below the
critical shear rate recently predicted (\Name{Denkov \etal}
\REVIEW{Phys. Rev. Lett.}{103}{2009}{118302}). Local measurements
of the constitutive law finally show that these foams behave as
simple Herschel-Bulkley yield stress fluids.}

\begin{document}

\maketitle

{\bf Introduction --} Materials such as dense suspensions,
colloidal gels, concentrated emulsions, foams or granular
materials, present a jammed structure \cite{LiuNagel2001}. This
results in the existence of a yield stress $\tau_y$ below which
they cannot flow. It was recently shown that some of these
materials exhibit shear banding while others do not
\cite{Dennin2008,Ovarlez2009,Schall2010}. Shear banding means that
in some conditions (near $\tau_y$), in a homogeneous shear stress
field, there is coexistence of liquid (sheared) and solid
(unsheared) regions \cite{Coussot2002b,Moller2008,Ovarlez2009}. At
the interface between both regions, the material flows at a
non-zero critical shear rate\footnote{The situation is different
in heterogeneous stress fields where shear localization occurs
with any yield stress fluid. The shear rate at the interface
between the sheared and unsheared regions is then naught for
simple, non shear banding, yield stress fluids, whereas it is
non-zero for shear banding materials \cite{Ovarlez2009}.}
$\gdot_c$. This implies that shear banding materials cannot flow
steadily at a shear rate smaller than $\gdot_c$. If one shears
such systems between two boundaries at a low macroscopic shear
rate $\gdot_{macro}<\gdot_c$, local velocity measurements
\cite{Manneville2008} then reveal shear banding
\cite{Ovarlez2009}: the material splits into a region flowing at
$\gdot_c$ and a non flowing region, whose relative extent ensures
that the shear rate spatial average is equal to $\gdot_{macro}$.
In macroscopic rheometric experiments, under applied stress, this
leads to the viscosity bifurcation phenomenon
\cite{Coussot2002a,DaCruz2002}. This implies that, in addition to
their yield stress, shear banding materials are also characterized
by a critical shear rate $\gdot_c$, thus defining a timescale
which must have its origin at a microscopic scale. Therefore, two
different classes of jammed systems should be considered depending
on whether they exhibit shear banding or not. To understand the
physical origin of this phenomenon, it is thus of high importance
to identify which materials do exhibit shear banding and which do
not.

Shear banding of jammed systems is observed mostly in colloidal
gels \cite{Coussot2002b,Moller2008,Rogers2008}, where aggregation
due to attractive interactions and thermally activated
structuration mechanisms is in competition with shear. It is also
observed in suspensions of noncolloidal particles where it results
from competition between sedimentation-induced contact formation
(which can be seen as an attractive force) and shear-induced
resuspension \cite{Fall2009}. Some yield stress fluids seem not to
exhibit shear banding, although experiments only provide upper
bounds on $\gdot_c$: neutrally buoyant noncolloidal suspensions
\cite{Fall2009}, Carbopol gels \cite{Coussot2009}, dense emulsions
\cite{Ovarlez2008}. In all these cases, if a critical shear rate
exists, it should be smaller than $10^{-2}$~s$^{-1}$. Hence, it is
suggested that attractive, thixotropic systems tend to develop
shear banding whereas repulsive and non-thermal systems do not
\cite{Rogers2008,Ovarlez2009,Coussot2009}.

In this context, the case of foams is rather puzzling. Disordered
foams and dense emulsions are both soft jammed packings of two
immiscible fluids with similar structures and mechanical behaviors
\cite{Hohler2005}. Their solid-like properties arise from their
density of interfacial energy, and steady flow is accomplished by
shear-induced structural rearrangements. Therefore, one could
\textit{a priori} expect foams to behave like emulsions with
respect to shear banding. Shear banding in emulsions was reported
once by B\'ecu \etal\ \cite{Becu2006} but this result was later
contradicted by Ovarlez \etal\ \cite{Ovarlez2008} who found
$\gdot_c<10^{-2}$~s$^{-1}$ for several emulsions. It is now
believed that the shear bands observed in \cite{Becu2006} are
transient shear bands (as observed in Carbopol gels
\cite{Divoux2010}), which disappear in steady state. Several cases
of shear banding in 2D and 3D foams have been reported in the
literature. Many observations of shear banding in 2D foams are now
attributed to viscous damping at the glass boundary
\cite{Wang2006,Janiaud2006,Krishan2008,Katgert2010} when the
bubbles are confined by one (or two) glass plate. The case of
bubble rafts is still unclear. On the one hand, Gilbreth \etal\
\cite{Gilbreth2006} have reported a consistent set of shear banded
velocity profiles characterized by a unique $\gdot_c$; these
profiles are claimed to be consistent with a single continuum
model in a 10 to 20 bubbles-wide zone, although no stress
measurement is reported. On the other hand, Katgert \etal\
\cite{Katgert2010} do not observe any signature of shear banding
for rafts of equivalent bubble size to gap ratios. Moreover their
velocity profiles are not consistent with a single constitutive
law, but can be described by a non-local model, such as what is
found in confined emulsions \cite{Goyon2008}.

The only measurement of local bulk velocity profile we are aware
of in steadily sheared 3D foams was performed by Rodts \etal\
\cite{Rodts2005} and revealed strong shear banding. Velocity
profiles were measured by MRI in a wet foam (gas volume fraction
$\phi=92\%$) sheared in a Couette geometry. At low rotational
velocity of the inner cylinder, flow localization was observed
with a discontinuity of the shear rate at the interface between
the sheared and the unsheared regions, thus providing a critical
shear rate $\gdot_c\simeq5$ s$^{-1}$. Consistently, macroscopic
strain rate measurements by Da Cruz \etal\ \cite{DaCruz2002} in a
similar 3D foam sheared in a homogeneous stress field (cone and
plate geometry) showed a viscosity bifurcation, characterized by a
critical shear rate $\gdot_c\simeq10$ s$^{-1}$.

\begin{figure*}\begin{center}
\includegraphics[width=5.9cm]{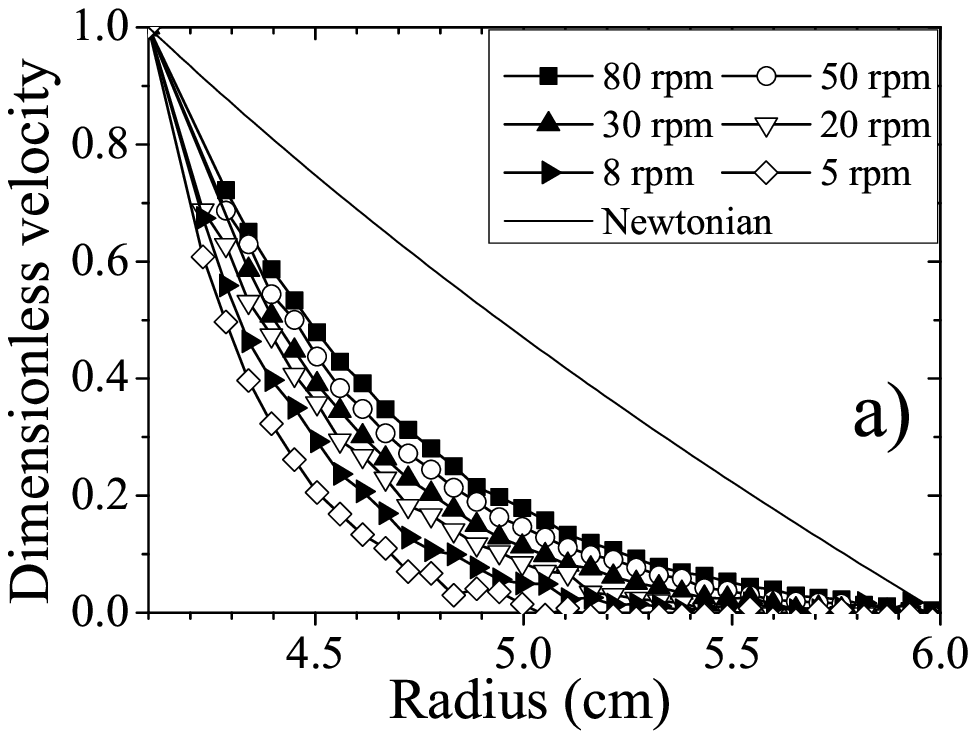}
\includegraphics[width=5.9cm]{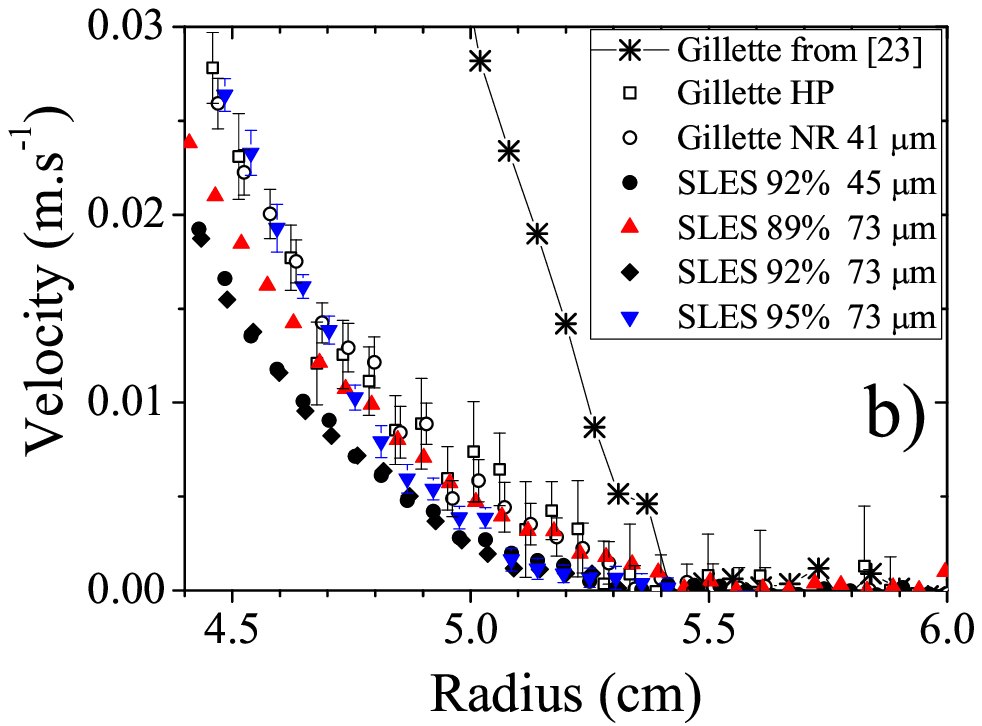}
\includegraphics[width=5.9cm]{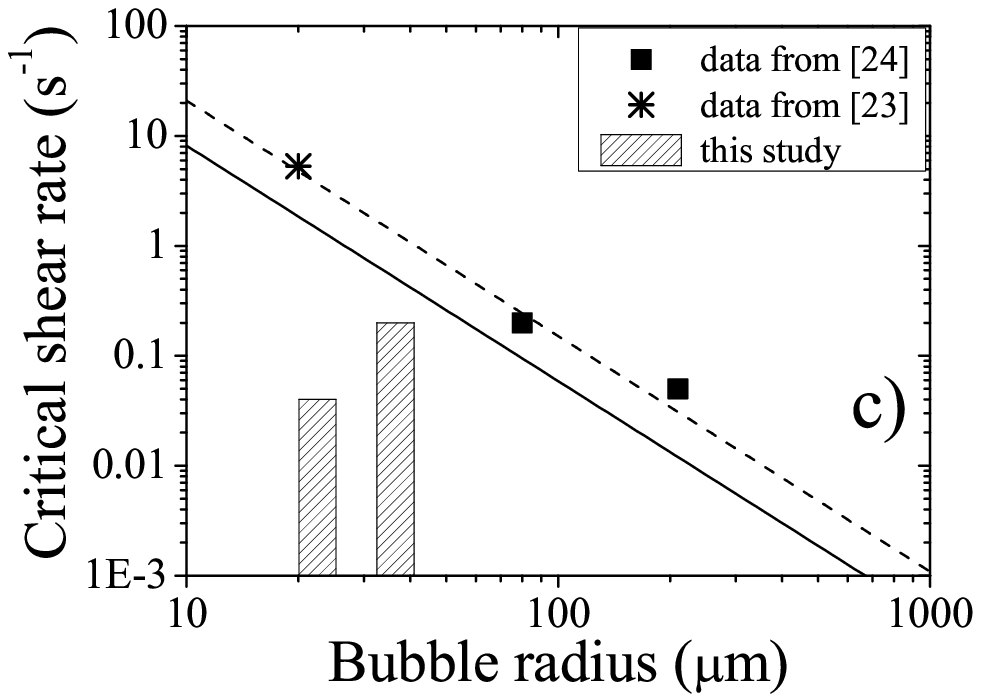}
\caption{a) Dimensionless velocity profiles $V(R)/V(R_i)$ for the
steady flows of SLES foam ($\phi=92\%$, $\left<d\right>=45\
\mu$m), at various rotational velocities ranging from 5 to 80~rpm;
the solid line is the theoretical profile for a Newtonian fluid.
b) Comparison between localized velocity profiles of the foams we
studied and that of Rodts \etal\ \cite{Rodts2005} (see $\phi$ and
$\left<d\right>$ in legend). c) Critical shear rate vs. bubble
radius for foams with $\phi=92$-$93\%$: Denkov \etal\ data
\cite{DenkovPRL2009} (squares), Rodts \etal\ data \cite{Rodts2005}
(star); the hatched bars show the intervals of possible critical
shear rates -- if any -- for the SLES foams in this study; the
lines are Eq.~\ref{equation_denkov} with $A_H=4\,10^{-20}$ J
(dashed line: $T=22.3$ mN/m, $\phi=93\%$, $\eta=3.8$ mPa\,s; solid
line: $T=31.0$ mN/m, $\phi=92\%$, $\eta=10.8$ mPa\,s).}
\label{fig1}
\end{center} \end{figure*}

To account for the observed shear banding in 3D foams and predict
the value of $\gdot_c$, Denkov \etal\ have proposed a model based
on the thinning dynamics of the films between bubbles
\cite{DenkovPRL2009}. Shear flow induces rearrangements thus
renewing bubble contacts. New films thin due to the capillary
pressure set by the curvature of the Plateau borders. As their
thickness reaches the range of van der Waals interactions, the
film may thin abruptly to a Newton black film (of a few nm
thickness) providing a strong adhesion between the bubbles and
locally jamming the bubble packing. A critical shear rate then
results from the competition between the timescale of
shear-induced rearrangements and that of the Newton black film
formation. It reads: \bea \gdot_c=
1.9\frac{T^{3/7}A_H^{4/7}}{\eta\,
{\left<d\right>}^{15/7}\left(1-\phi\right)^{0.3}}\label{equation_denkov}\eea
with $A_H$ the Hamaker constant ($A_H\simeq4\,10^{-20}$ J), $T$
the surface tension, $\eta$ the viscosity of the solution, and
$\left<d\right>$ the average bubble diameter. For a given $\phi$,
$\gdot_c$ does not vary much with $T$ and exhibits a marked
dependency with $\eta$ and $\left<d\right>$. For instance, the
value of $\gdot_c$ expected for a typical Gillette foam is 10
s$^{-1}$ (with $T$=28.6 mN/m, $\eta$= 1.9 mPas, $\phi$=92\% and
$\left<d\right>$= 40 $\mu$m). This prediction (see
Fig.~\ref{fig1}c) seems to be in good agreement with MRI data
\cite{Rodts2005} and with recent qualitative measurements
\cite{DenkovPRL2009}.

Finally, these results tend to show that 3D foams are not simple
yield stress fluids, in contrast with dense emulsions. However,
whereas solely local measurements of bulk velocity profiles can
unambiguously determine the existence of steady shear banding in
3D foams, only one such measurement exists in the literature
\cite{Rodts2005}. In this letter, we present a set of new
experiments in 3D foams, with different gas volume fractions,
bubble size and interfacial rheological properties. Local velocity
profiles measured using MRI show that, in contrast with
\cite{DaCruz2002,Rodts2005}, 3D foams do not exhibit any
observable shear banding. Consistently, they do not exhibit
viscosity bifurcation. They appear to behave as simple
Herschel-Bulkley fluids, as previously found in emulsions
\cite{Ovarlez2008}.

{\bf Materials and methods --} We study a foam based on a mixed
surfactant aqueous solution, denoted SLES. It contains Sodium
Lauryl-dioxyethylene Sulfate (Stepan Co., USA) with concentration
0.33\% g/g, Cocoamidopropyl Betaine (Goldschmidt, Germany) with
concentration 0.17\% g/g and 60\% g/g Glycerol (Fluka, anhydrous
p.a. 99.5\% GC). The chemicals are dissolved in water (millipore
milli-Q). The surface tension and the viscosity of this solution
at $21\pm1$\degre$\!$C are $T=31.0$ mN/m and $\eta=10.8$ mPa\,s
\cite{Krishan2010}. Foam is produced by simultaneously flowing the
SLES solution and pressurized nitrogen gas saturated with
perfluorohexane vapor through a glass bead column
\cite{Rouyer2003}. By adjusting the liquid flow rate and gas
pressure, we produce samples with different controlled gas volume
fractions $\phi$. Using videomicroscopy, we characterize the
average $\left<d\right>$ of the bubble diameter $d$ distribution
and its dimensionless standard deviation
$\mu=\sqrt{\left<d^2\right>-\left<d\right>^2}/\left<d\right>$ for
each foam sample. We prepare three materials with same
$\left<d\right>=73\,\mu$m (and $\mu$ comprised between 0.5 and
0.6), of different $\phi$: (i) $(88.4\pm0.3)\%$, (ii)
$(92.5\pm0.3)\%$ , and (iii) $(95.3\pm0.3)\%$. In addition, we
have a sample with $\left<d\right>=45\,\mu$m, $\mu=0.7$, and
$\phi=(92.5\pm0.3)\%$. We also use Gillette shaving cream (Normal
Regular), denoted Gillette NR, as in previous studies
\cite{Gopal1999} for the sake of comparison with
\cite{DaCruz2002}, with $\phi=(92.0\pm0.5)\%$,
$\left<d\right>=41\,\mu$m and $\mu=0.6$. Its constitutive foaming
solution has T = 28.6 mN/m and $\eta=1.9$ mPa\,s
\cite{Krishan2010}. For the sake of comparison with
\cite{Rodts2005}, we use another Gillette shaving cream (Haute
Protection) denoted Gillette HP. As the rigidity of the liquid-gas
interfaces plays a role on the behavior of steadily sheared foams
\cite{DenkovSoftMatt2009}, it is interesting to test the influence
of this parameter on shear banding. This can be done from our
experiments (at $\phi=92\%$ and $\left<d\right>\simeq40\mu$m)
since in contrast to SLES foam, Gillette NR foam has rigid
liquid-gas interfaces \cite{DenkovSoftMatt2009,Krishan2010}.

The material's local behavior is studied in a wide-gap Couette
geometry (inner cylinder radius $R_i$=4.1~cm; outer cylinder
radius $R_o$=6~cm; height $H$=11~cm). The use of a wide gap likely
prevents (non-local) finite size effects
\cite{Goyon2008,Ovarlez2008,Katgert2010}. Sandpaper is glued to
the walls to avoid slippage; there is no observable slip in the
velocity profiles. The rheometer is inserted in a 0.5-T vertical
MRI spectrometer (24/80 DBX by Bruker). In all experiments, the
velocity $\Omega$ of the inner cylinder is controlled. We measure
the local velocity in the flowing sample for various constant
$\Omega$ ranging from 5 to 100~rpm; the torque exerted by the
material on the inner cylinder is measured using a Bohlin C-VOR
200 rheometer. The orthoradial velocity profiles $V_\theta(R)$ are
obtained through MRI techniques as described in
\cite{Rodts2004,Ovarlez2006}. The MRI setup also allows
measurement of the local water content \cite{Ovarlez2006}. In all
experiments, we checked that drainage can be neglected (it led to
absolute volume fraction variations of order 0.5\% in the 4cm high
measurement zone over the duration of the experiments). We also
checked that there is no observable shear-induced radial migration
of water, as in emulsions \cite{Ovarlez2008}.

As a complement and to allow for a direct comparison of our data
with previous ones \cite{DaCruz2002}, we have performed
macroscopic viscosity bifurcation experiments. Such macroscopic
experiments are convincing only if the stress field is
homogeneous; we thus use a 4\!\degre cone and plate geometry (as
in \cite{DaCruz2002}), with serrated surfaces of 30~mm radius; the
stress heterogeneity is then of the order of 0.5\%. Note however
that the gap is small (the gap at the edge is 50 bubbles high);
therefore, one cannot exclude finite size (non-local) effects
\cite{Goyon2008,Katgert2010}. Experiments consist in first
applying a preshear at high shear rate (20~s$^{-1}$) during 60~s,
ensuring that the whole material flows initially, before applying
a constant stress for 150 to 300~s. The macroscopic shear rate is
then plotted vs. time to check for the existence of a steady flow.

{\bf Velocity profiles and shear banding --} In this section, we
focus on the stationary velocity profiles. In Fig.~\ref{fig1}a, we
plot the dimensionless velocity profiles $V(R)/V(R_i)$ obtained
with SLES foam ($\phi=92\%$) for various rotational velocities
$\Omega$. All the studied foams exhibit similar behavior. We
observe that the material is sheared only in a fraction of the gap
at low $\Omega$: $V(R)$ vanishes (within the measurement
uncertainty) at some radius $R_c(\Omega)<R_o$. $R_c$ increases
with $\Omega$. Beyond a critical velocity (that depends on the
foam and is of order 40~rpm in Fig.~\ref{fig1}a), the whole sample
is sheared.

To analyze these observations, one should first be reminded that
the shear stress distribution $\tau(R)$ is heterogeneous: it
decreases with increasing radius $R$ and reads
$\tau(R)=\tau(R_i)\,R_i^2/R^2$ from stress balance equations. When
shear extends over the whole gap, the velocity profiles differ
from those of a Newtonian fluid. Their strong curvature is due to
the stress heterogeneity (since $\tau(R_i)/\tau(R_o)=2.1$) and is
typical of shear-thinning fluid: the shear rate decreases more
rapidly within the gap (\textit{i.e.} when $\tau$ decreases) than
for a Newtonian fluid. The shear localization observed at low
velocity is a feature of yield stress fluid flow in Couette
geometry. At low $\Omega$ (or applied stress slightly above the
material yield stress $\tau_y$), the flow stops at a radius $R_c$
within the gap where the local shear stress $\tau(R)$ equals
$\tau_y$ (\textit{i.e.} $R_c=R_i\sqrt{\tau(R_i)/\tau_y}$). The
decrease of $R_c(\Omega)$ as $\Omega$ is decreased then comes from
the rate dependence of the constitutive law at the approach of
$\tau_y$. Note the discrepancy between this observation and that
of Katgert \etal\ in 2D foams \cite{Katgert2010} where
dimensionless velocity profiles superpose and extend over the
whole gap at any low $\Omega$. We will discuss this differences in
the section devoted to the constitutive law measurements.

From the shear localized velocity profiles, we now analyze the
possible shear banding behavior of the studied foams. Two cases
should be considered. On the one hand, if the material is a shear
banding material, the shear rate $\gdot(R)$ should tend towards a
critical shear rate $\gdot_c\neq0$ as $\tau$ approaches $\tau_y$
\textit{i.e.} as $R$ tends to $R_c$. As $\gdot(R)=V/R-\upd V/\upd
R$, the local velocity should tend to zero with a non-zero slope
$|\upd V/\upd R|=\gdot_c$, independent of the velocity at the
inner cylinder. Shear banding should then result in a
discontinuity of the slope of the velocity profile at the
interface between the sheared and the unsheared regions, as in a
homogeneous stress field \cite{Ovarlez2009}. This is precisely as
observed by Rodts \etal\ \cite{Rodts2005} (see Fig.~\ref{fig1}b).
For various $\Omega$, they found a consistent set of shear banded
velocity profiles with a slope $\gdot_c\approx$5~s$^{-1}$ at the
interface. On the other hand, if the material is not a shear
banding material, $\gdot(R)$ should tend continuously to zero as
$\tau$ approaches $\tau_y$ \textit{i.e.} as $R$ tends to $R_c$.
This would mean that $\upd V/\upd R=0$ at the interface between
the sheared and the unsheared regions \textit{i.e.} the velocity
profile should tend smoothly to zero.

In Fig.~\ref{fig1}b, we plot shear localized velocity profiles
obtained by Rodts \etal\ and by us corresponding to a same
position of $R_c$ (here $\simeq5.4$~cm). We focus on the shape of
the velocity profiles at the interface between the sheared and the
unsheared regions. From this plot, it is clear that, in contrast
with the Rodts \etal\ data, all of our systems exhibit a smooth
transition from flow to rest and thus seem to be not shear banding
materials. Of course, due to finite experimental resolution, we
can never be certain that no critical shear rate exists. We can
only provide upper bounds on the critical shear rate -- if any --
as shown in Tab.~\ref{table1}. Note that the accuracy of the shear
rate measurement depends on the spatial resolution and on the MRI
signal to noise ratio, which depends itself on the foaming
solution and on external factors. The best experimental conditions
were met with the 92\% SLES foam, and led to a rather low upper
bound $\approx 0.04$~s$^{-1}$.

\begin{table}[htbp] \begin{center}\begin{tabular}{ccccc} Foam&$\phi$&$\left<d\right>$&Upper bound&Model\\
& & & on $\gdot_c$ &(Eq.~\ref{equation_denkov})\\ \hline
SLES& 92\% & 45~$\mu$m&0.04~s$^{-1}$&1.5~s$^{-1}$\\
SLES& 88\% & 73~$\mu$m&0.2~s$^{-1}$&0.45~s$^{-1}$\\
SLES& 92\% & 73~$\mu$m&0.2~s$^{-1}$&0.52~s$^{-1}$\\
SLES& 95\% & 73~$\mu$m&0.3~s$^{-1}$&0.60~s$^{-1}$\\
Gillette NR& 92\% & 41~$\mu$m &0.2~s$^{-1}$ & 9.6~s$^{-1}$\\
Gillette HP& -- & -- &0.6~s$^{-1}$&--
\end{tabular}\caption{Upper bounds on the critical shear rate $\gdot_c$ obtained from MRI measurements and predictions of Eq.~\ref{equation_denkov} for all foams.}
\label{table1}\end{center}\end{table}

\textbf{Viscosity bifurcation --} It is striking that we find
upper bounds much lower than the critical shear rates $\gdot_c$
previously reported \cite{DaCruz2002,Rodts2005}. From macroscopic
measurements, Da Cruz \etal\ evidenced a viscosity bifurcation in
the Gillette NR foam \cite{DaCruz2002}, with
$\gdot_c\approx$10~s$^{-1}$. To understand if viscosity
bifurcation experiments can be reconciled with our local
measurements, we have performed these experiments on some of our
systems, including the Gillette NR foam of \cite{DaCruz2002}. We
focus on measurements obtained for applied stress close to the
yield stress to estimate the lowest steady state shear rates than
can be achieved.

\begin{figure}\begin{center}
\includegraphics[width=5.8cm]{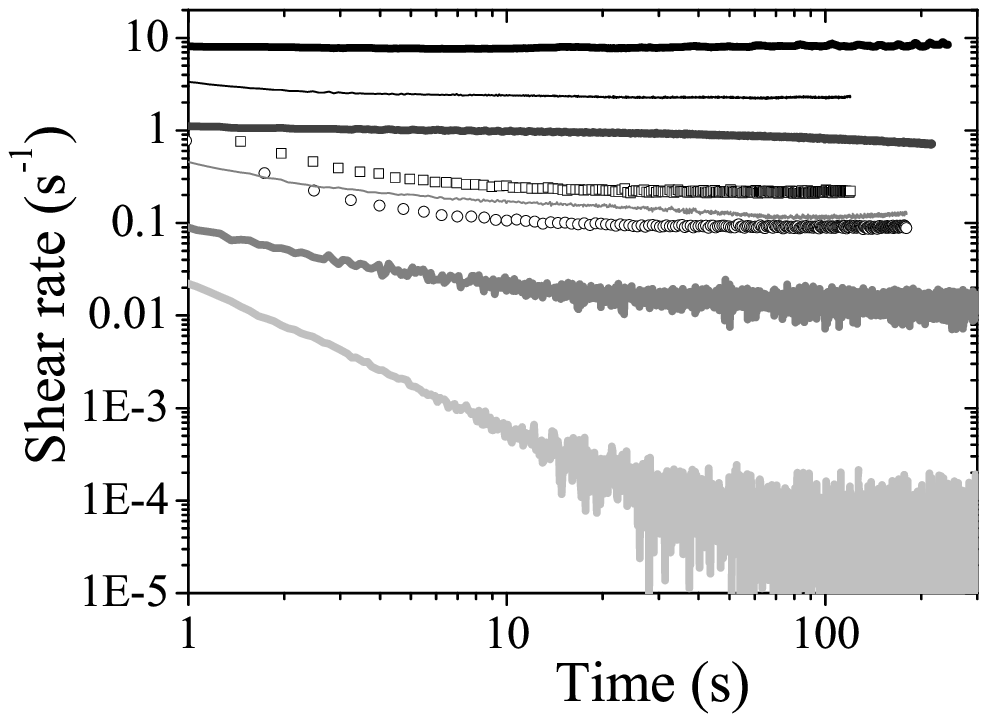}
\caption{Viscosity bifurcation experiments: shear rate vs. time at
constant controlled stress in a cone and plate geometry. Thick
lines: SLES foam with $\phi=92\%$ and $\left<d\right>=45\mu$m (in
grey scale, from black to light grey: 65 Pa, 45 Pa, 30 Pa, 20 Pa).
Thin lines: Gillette HP as in \cite{Rodts2005} (black: 60 Pa,
grey: 35 Pa). Empty symbols: Gillette NR as in \cite{DaCruz2002},
with $\left<d\right>=30\mu$m (squares: 60 Pa, circles: 50
Pa).}\label{fig2}
\end{center} \end{figure}

Fig.~\ref{fig2} shows that steady state flow can be achieved under
controlled stress conditions at shear rate values as low as
0.1~s$^{-1}$ for the two Gillette foams and 0.02~s$^{-1}$ for the
SLES foam. Further experiments would have to be performed to
improve the estimate of the upper bound of $\gdot_c$ from these
indirect measurements. At this stage, the conclusion is that local
and macroscopic measurements are consistent, and lead to the same
result: the absence of observable shear banding in all foams we
studied.

\textbf{Analysis --} We now question the possible origin of the
differences between our results and the previous ones, obtained on
the same systems (the Gillette foams of
\cite{DaCruz2002,Rodts2005}) using the same experimental displays.
 First, Da Cruz \etal\ \cite{DaCruz2002} report a 200~s$^{-1}$
preshear; we observe that applying such high shear rate to the
Gillette NR foam with the geometry used in \cite{DaCruz2002} leads
to expelling a large amount of the material from the gap by
inertial forces (this happens for shear rates higher than
30~s$^{-1}$); therefore the protocol was inappropriate, leading to
incorrect rheological data. Then the yield stress reported is
$\approx$180 Pa, a value much higher than values reported in the
literature \cite{Hohler2005,Gopal1999,Rouyer2005}. This would mean
that the material on which the measurements were performed in
\cite{DaCruz2002} was not what it was supposed to be. The case of
the Rodts \etal\ data \cite{Rodts2005} is less obvious. They
report a 7 Pa yield stress, which is lower than data from the
literature \cite{Hohler2005,Gopal1999,Rouyer2005}. Such a low
value may indicate a material that is wetter than it should be,
yet it is unclear why this would lead to shear banding. It was
recently shown that some non-shear-banding materials may exhibit
transient shear banding \cite{Divoux2010} when a low velocity is
applied after a short resting period; this is not observed when
the low velocity is applied just after a high velocity. In order
to check that the observations of Rodts \etal\ were not due to
such a transient effect, we have studied the role of the shear
history. MRI velocity measurements were performed under various
conditions, when going from high to low velocities, from low to
high velocities, and when rotating directly after a long resting
time: all procedures were found to provide the same velocity
profiles, \textit{i.e.} no shear banding (even transient) was
observed. The occurrence of shear banding in \cite{Rodts2005}
might be due to uncontrolled traces of impurities in the system,
e.g. clay particles, which were also studied with the same
equipment. It was indeed shown that tiny amounts of colloidal clay
particles dispersed in the continuous phase of a simple emulsion
could form bridges between droplets and lead to thixotropic
effects and shear banding \cite{Ragouilliaux2007}; the same might
happen with a foam. Finally, let us stress that we provide here a
consistent set of data performed with five different foams,
including the Gillette foams of \cite{DaCruz2002,Rodts2005}, with
two different experiments. The conclusion is the absence of
observable shear banding in all the foams we studied.

We now compare our upper bounds on $\gdot_c$ with the model of
Denkov \etal\ \cite{DenkovPRL2009} (Eq.~\ref{equation_denkov}).
Fig.~\ref{fig1}c shows these bounds vs. the bubble radius for SLES
foam ($\phi=92\%$) as well as the $\gdot_c$ value reported in
\cite{Rodts2005}. It also shows the data obtained by Denkov \etal\
with foams of the same surfactants as in SLES foam mixed with
myristic acid. Note that these last data were obtained in a 8 to
15 bubbles-wide gap, a case where finite size effects may be
observed \cite{Katgert2010}. When the timescale of shear $1/\gdot$
is lower than the duration of bubble rearrangements, there may be
stress heterogeneities at the bubble scale, leading to shear
banding in a thin layer \cite{Kabla2007}, which is not bulk shear
banding.

There is a strong discrepancy between the model and our data; the
model overestimates the critical shear rate $\gdot_c$ for all the
studied foams, as shown in Tab.~\ref{table1}. This means that,
while the proposed mechanism in \cite{DenkovPRL2009} is certainly
relevant in some cases, it may not apply to the foams made from
ionic surfactant solutions without salt addition that are used in
this study. Liquid films actually thin until the balance between
capillary and disjoining pressures is reached, and form
equilibrium common black films of thickness in the range 10-100 nm
rather than very thin Newton black films. Moreover, the model
assumes rigid interfaces, which sets the boundary conditions of
the film thinning flow profile. However mobile interfaces, as in
SLES foam, should lead to faster thinning, and thus to even higher
critical shear rate than with rigid interfaces. This qualitative
argument shows that the interfacial rigidity cannot account for
the difference between the experimental upper bound of $\gdot_c$
and Eq.~\ref{equation_denkov}. More experiments, in systems with
strong attraction between the films, are needed to further test
the model.

Nevertheless, as noted in the introduction, the absence of shear
banding in foams is consistent with the basic mechanisms that tune
shear banding in other complex fluids \cite{Ovarlez2009}. In
particular, it is consistent with observations in dense emulsions
\cite{Ovarlez2008}.

\begin{figure}\begin{center}
\includegraphics[width=5.85cm]{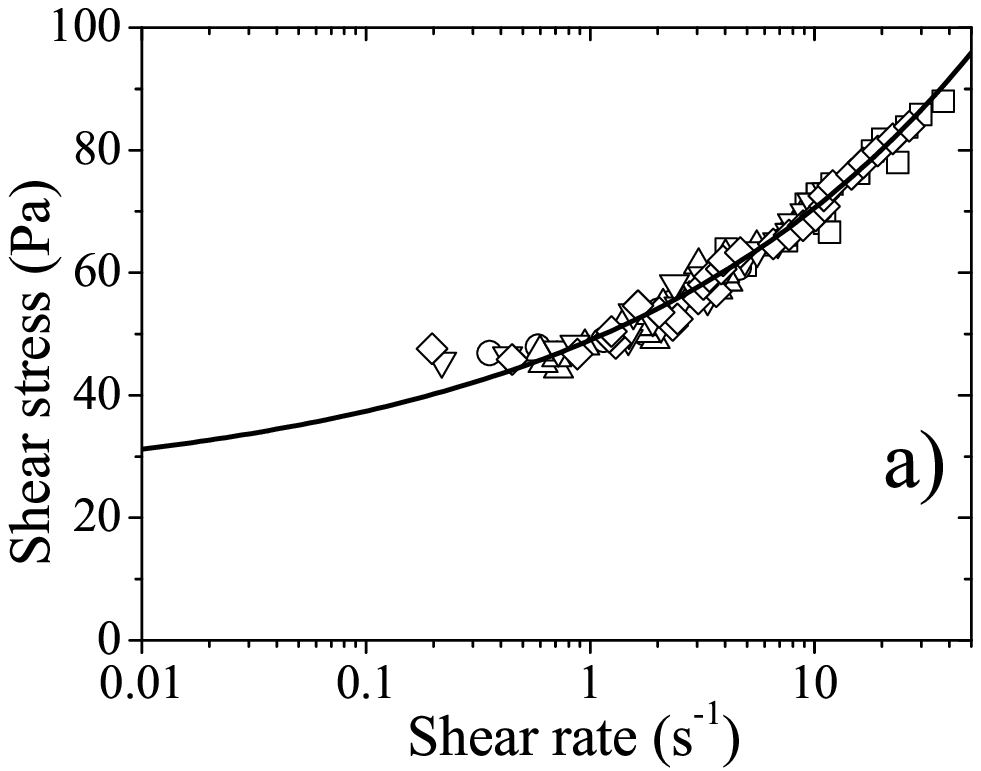}
\includegraphics[width=5.85cm]{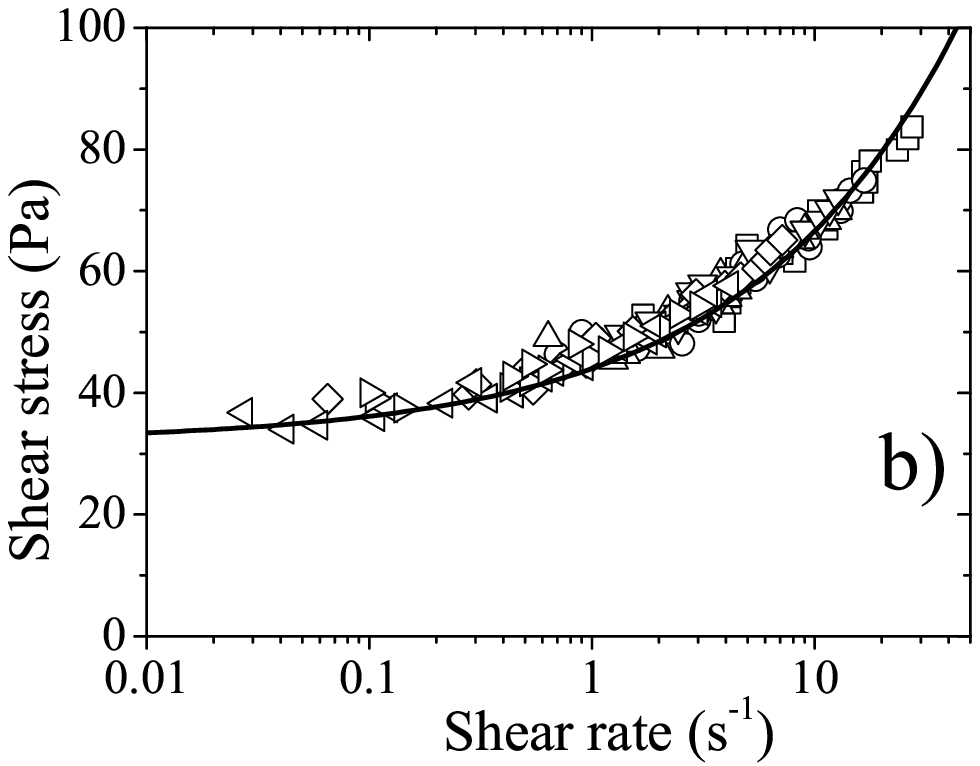}
\caption{Local constitutive laws measurements of a) Gillette NR
foam ($\left<d\right>=41\,\mu$m), and b) SLES foam ($\phi=92\%$,
$\left<d\right>=45\,\mu$m). Each symbol corresponds to local
measurements performed at a same rotational velocity. The solid
lines are Herschel-Bulkley fits to the data
$\tau=\tau_y+\eta_{HB}\gdot^n$, with $\tau_y=24$Pa,
$\eta_{HB}=25$Pa.s$^{n}$, $n=0.27$ for the Gillette NR foam,
$\tau_y=32$Pa, $\eta_{HB}=12$Pa.s$^{n}$, $n=0.46$ for the SLES
foam.}\label{fig3}
\end{center} \end{figure}

{\bf Constitutive law --} The constitutive laws accounting for the
materials' velocity profiles can be built from our experimental
data. From the torque measurements $T$ and the stress balance
equations, one gets the shear stress distribution in the gap
$\tau(R)=T/(2\pi HR^2)$. The local shear rate $\gdot(R)$ in the
gap is inferred from the velocity profiles $V(R)$ through
$\gdot(R)=V/R-\upd V/\upd R$. Both measurements performed at a
given radius $R$ for a given rotational velocity $\Omega$ thus
provide a data point of the local constitutive law. We checked
that the materials remain homogeneous unpon shear, which allows us
to combine the data measured at various radii. The local
constitutive laws $\tau(\gdot)$ obtained for two of the studied
foams are plotted in Fig.~\ref{fig3}a~and~\ref{fig3}b.

We find that the local flows curves obtained from experiments
performed at various $\Omega$ are consistent with a single --
local -- constitutive law for each foam. This contrasts with
recent results in 2D sheared foams \cite{Katgert2010} where no
single local constitutive law can account for flow profiles
obtained at different $\Omega$. The same difference -- non-local
and local constitutive laws -- is observed with emulsions flows in
microchannels \cite{Goyon2008} or in our wide gap Couette geometry
\cite{Ovarlez2008}. As in \cite{Goyon2008,Ovarlez2008}, the main
difference between our observations and those of Katgert \etal\
\cite{Katgert2010} may stand in the bubble size to gap ratio
(\textit{i.e.} nonlocal effects are finite size effects). The gap
of our geometry is indeed about 500 bubbles wide, whereas the gaps
used in \cite{Katgert2010} are 20 to 40 bubbles wide. Leaving
these possible finite size effects apart, it can finally be
concluded that foams, together with emulsions \cite{Ovarlez2008}
and Carbopol gels \cite{Coussot2009}, belong to the class of
simple yield stress fluids \cite{Ovarlez2009}.

The local constitutive laws we obtain are well fitted to
Herschel-Bulkley laws, with exponents consistent with those of
foams with rigid (n=0.27) or mobile (n=0.46) interfaces
\cite{DenkovSoftMatt2009}. We also remark that the
Herschel-Bulkley consistency $\eta_{HB}$ of the SLES foam is in
fair agreement with the Denkov \etal\ prediction (see Eq.~5 of
\cite{DenkovSoftMatt2009}). Note finally that, due to the limited
range of low strain rates measurements, the yield stress for
Gillette NR foam is probably underestimated

{\bf Conclusion --}  Our measurements demonstrate that
three-dimensional foams do not exhibit observable signatures of
shear banding, in contrast with previous observations
\cite{DaCruz2002,Rodts2005}. The question remains open whether
steady shear banding may occur in extremely dry foams since it has
been observed upon shear start-up \cite{Rouyer2003}. Our results
disagree with the recent model of Denkov \etal\
\cite{DenkovPRL2009}, which is shown here to overestimate the
critical shear rate -- if any -- of the foams we studied. Further
experiments with strong attraction between films would be needed
to test the mechanism proposed in \cite{DenkovPRL2009}. While
nonlocal effects have been recently evidenced \cite{Katgert2010},
we have finally shown that the constitutive law of foams measured
locally in a wide gap geometry is that of simple Herschel-Bulkley
yield stress fluids, as emulsions \cite{Ovarlez2008} and Carbopol
gels \cite{Coussot2009}. Experiments at both small and large
scales on a same system, as in of Goyon \etal\ \cite{Goyon2008},
remain to be performed to test the non-local modelling of 3D foam
flows.

\acknowledgments We thank P.~Coussot and S.~Rodts for providing
their data and for open discussions on their papers. We also thank
R.~H\"ohler and N.~Denkov for fruitful discussions, and H.~Sizun
for crucial technical help. We gratefully acknowledge financial
support from E.S.A. (MAP No.~AO99-108: C14914/02/NL/SH).


\begin{thebibliography}{0}

\bibitem{LiuNagel2001} \Editor{Liu A.~J. \and
Nagel S.~R.} \Book{Jamming and Rheology: Constrained Dynamics on
Microscopic and Macroscopic Scales} \Publ{Taylor \& Francis, New
York} \Year{2001}.

\bibitem{Dennin2008} \Name{Dennin M.} \REVIEW{J. Phys. Condens.
Matter}{20}{2008}{283103}.

\bibitem{Ovarlez2009}\Name{Ovarlez G., Rodts S., Chateau X. \and Coussot P.}
\REVIEW{Rheol. Acta}{48}{2009}{831}.

\bibitem{Schall2010} \Name{Schall P. \and
van Hecke M.} \REVIEW{Annu. Rev. Fluid Mech.}{42}{2010}{67}.

\bibitem{Coussot2002b} \Name{Coussot P., Raynaud J.~S., Bertrand F.,
Moucheront P., Guilbaud J.~P., Huynh H.~T., Jarny S. \and Lesueur
D.} \REVIEW{Phys. Rev. Lett.}{88}{2002}{218301}.

\bibitem{Moller2008} \Name{M\o ller P.~C.~F., Rodts S., Michels M.~A.~J. \and Bonn D.}
\REVIEW{Phys. Rev. E}{77}{2008}{041507}.

\bibitem{Manneville2008}\Name{Manneville S.}
\REVIEW{Rheol. Acta}{47}{2008}{301}.

\bibitem{Coussot2002a}
\Name{Coussot P., Nguyen Q.~D., Huynh H.~T. \and Bonn D.}
\REVIEW{Phys. Rev. Lett.}{88}{2002}{175501}.

\bibitem{DaCruz2002}
\Name{Da Cruz F., Chevoir F., Bonn D. \and Coussot P.}
\REVIEW{Phys. Rev. E}{66}{2002}{051305}.

\bibitem{Rogers2008}\Name{Rogers S.~A., Vlassopoulos D. \and
Callaghan P.~T.} \REVIEW{Phys. Rev. Lett.}{100}{2008}{128304}.

\bibitem{Fall2009} \Name{Fall A., Bertrand F., Ovarlez
G. \and Bonn D.} \REVIEW{Phys. Rev. Lett.}{103}{2009}{178301}.

\bibitem{Coussot2009} \Name{Coussot P., Tocquer L., Lanos C. \and Ovarlez G.}
\REVIEW{J. Non-Newtonian Fluid Mech.}{158}{2009}{85}.

\bibitem{Ovarlez2008}\Name{Ovarlez G., Rodts S.,
Coussot P., Goyon J. \and Colin A.} \REVIEW{Phys. Rev.
E}{78}{2008}{036307}.

\bibitem{Hohler2005}\Name{H\"ohler R. \and Cohen-Addad S.}
\REVIEW{J. Phys.: Condens. Matter}{17}{2005}{R1041}.

\bibitem{Becu2006} \Name{B\'ecu L., Manneville S. \and Colin A.} \REVIEW{Phys. Rev.
Lett.}{96}{2006}{138302}.

\bibitem{Divoux2010} \Name{Divoux T., Tamarii D., Barentin C. \and
Manneville S.} \REVIEW{Phys. Rev. Lett.}{104}{2010}{208301}.

\bibitem{Wang2006} \Name{Wang Y., Krishan K. \and Dennin M.} \REVIEW{Phys. Rev.
E}{73}{2006}{031401}.

\bibitem{Janiaud2006} \Name{Janiaud E., Weaire D. \and Hutzler S.}
\REVIEW{Phys. Rev. Lett.}{97}{2006}{038302}.

\bibitem{Krishan2008} \Name{Krishan K. \and Dennin M.} \REVIEW{Phys. Rev.
E}{78}{2008}{051504}.

\bibitem{Katgert2010} \Name{Katgert G.,
Tighe B.~P., M\"obius M.~E. \and van Hecke M.} \REVIEW{Europhys.
Lett.}{90}{2010}{54002}.

\bibitem{Gilbreth2006} \Name{Gilbreth C., Sullivan S. \and Dennin M.}
\REVIEW{Phys. Rev. E}{74}{2006}{051406}.

\bibitem{Goyon2008} \Name{Goyon J., Colin A., Ovarlez G., Ajdari
A. \and Bocquet L.} \REVIEW{Nature}{454}{2008}{84}.

\bibitem{Rodts2005}\Name{Rodts S., Baudez J.~C. \and Coussot P.}
\REVIEW{Europhys. Lett.}{69}{2005}{636}.

\bibitem{DenkovPRL2009} \Name{Denkov D., Tcholakova S.,
Golemanov K. \and Lips A.} \REVIEW{Phys. Rev.
Lett.}{103}{2009}{118302}.

\bibitem{Krishan2010} \Name{Krishan K., Helal A., H\"ohler R. \and Cohen-Addad S.}
\REVIEW{Phys. Rev. E}{82}{2010}{011405}.

\bibitem{Rouyer2003} \Name{Rouyer F., Cohen-Addad S., Vignes-Adler M. \and H\"ohler R.}
\REVIEW{Phys. Rev. E}{67}{2003}{021405}.

\bibitem{Gopal1999} \Name{Gopal A.~D. \and Durian D.~J.}
\REVIEW{Journal of Colloid and Interface Science}{213}{1999}{169}.

\bibitem{DenkovSoftMatt2009} \Name{Denkov N.D.,
Tcholakova S., Golemanov K., Ananthpadmanabhan K.P. \and Lips A.}
\REVIEW{Soft Matter}{5}{2009}{3389}.

\bibitem{Rodts2004}\Name{Rodts S., Bertrand F.,
Jarny S., Poullain P. \and Moucheront P.} \REVIEW{C. R.
Chim.}{7}{2004}{275}.

\bibitem{Ovarlez2006} \Name{Ovarlez G.,
Bertrand F. \and Rodts S.} \REVIEW{J. Rheol.}{50}{2006}{259}.

\bibitem{Rouyer2005} \Name{Rouyer F., Cohen-Addad S. \and H\"ohler R.} \REVIEW{Colloids
Surf. A}{263}{2005}{111}.

\bibitem{Ragouilliaux2007}
\Name{Ragouilliaux A., Ovarlez G., Shahidzadeh-Bonn N., Herzhaft
B., Palermo T. \and Coussot P.} \REVIEW{Phys. Rev.
E}{76}{2007}{051408}.

\bibitem{Kabla2007} \Name{Kabla A., Scheibert J. \and Debregeas G.}
\REVIEW{J. Fluid Mech.}{587}{2007}{45}.


\end{thebibliography}
\end{document}